\documentclass[a4paper,11pt]{article}
\usepackage{pos}

\usepackage{graphicx}
\usepackage{amsmath}
\usepackage{siunitx}

\title{Trigger Optimization and Event Classification for Dark Matter Searches in the CYGNO Experiment Using Machine Learning}
\ShortTitle{ML for DM searches in CYGNO}

\author[a]{F D Amaro}
\author[b,c]{R Antonietti}
\author[d,e]{E Baracchini}
\author[f]{L Benussi}
\author[f]{C Capoccia}
\author[f,g]{M Caponero}
\author[h]{L G M de Carvalho}
\author[i,j]{G Cavoto}
\author[f]{I A Costa}
\author[f]{A Croce}
\author[d,e]{M D'Astolfo}
\author[j]{G D'Imperio}
\author[f]{G Dho}
\author[j]{E Di Marco}
\author[a]{J M F dos Santos}
\author[d,e]{D Fiorina}
\author[j]{F Iacoangeli}
\author[d,e]{Z Islam}
\author[k]{E Kemp}
\author[d,e]{H P Lima Jr}
\author[f]{G Maccarrone}
\author[a]{R D P Mano}
\author[d,e]{D J G Marques}
\author[f]{G Mazzitelli}
\author[b,c]{P Meloni}
\author[i,j]{A Messina}
\author[a]{C M B Monteiro}
\author[h]{R A Nobrega}
\author*[d,e]{G M Oppedisano}  
\author[h]{I F Pains}
\author[f]{E Paoletti}
\author[b,c]{F Petrucci}
\author[d,e]{S Piacentini}
\author[f]{D Pierluigi}
\author[j]{D Pinci}
\author[j]{F Renga}
\author[f]{A Russo}
\author[f,l]{G Saviano}
\author[a]{P A O C Silva}
\author[m]{N J Spooner}
\author[f]{R Tesauro}
\author[f]{S Tomassini}
\author[i,j]{D Tozzi}

\affiliation[a]{LIBPhys, Department of Physics, University of Coimbra, 3004-516 Coimbra, Portugal}
\affiliation[b]{Dipartimento di Matematica e Fisica, Università Roma Tre, 00146 Roma, Italy}
\affiliation[c]{INFN Sezione di Roma Tre, 00146 Roma, Italy}
\affiliation[d]{Gran Sasso Science Institute, 67100 L'Aquila, Italy}
\affiliation[e]{INFN Laboratori Nazionali del Gran Sasso, 67100 Assergi, Italy}
\affiliation[f]{INFN Laboratori Nazionali di Frascati, 00044 Frascati, Italy}
\affiliation[g]{ENEA Centro Ricerche Frascati, 00044 Frascati, Italy}
\affiliation[h]{Universidade Federal de Juiz de Fora, Faculdade de Engenharia, 36036-900 Juiz de Fora, MG, Brazil}
\affiliation[i]{Dipartimento di Fisica, Sapienza Università di Roma, 00185 Roma, Italy}
\affiliation[j]{INFN Sezione di Roma, 00185 Roma, Italy}
\affiliation[k]{Universidade Estadual de Campinas (UNICAMP), Campinas 13083-859, SP, Brazil}
\affiliation[l]{Dipartimento di Ingegneria Chimica, Materiali e Ambiente, Sapienza Università di Roma, 00185 Roma, Italy}
\affiliation[m]{Department of Physics and Astronomy, University of Sheffield, Sheffield S3 7RH, UK}

\emailAdd{giuseppe.oppedisano@gssi.it}

\abstract{
The CYGNO experiment employs an optical-readout Time Projection Chamber (TPC) to search for rare low-energy interactions using finely resolved scintillation images. While the optical readout provides rich topological information, it produces large, sparse megapixel images that challenge real-time triggering, data reduction, and background discrimination.

We summarize two complementary machine-learning approaches developed within CYGNO. First, we present a fast and fully unsupervised strategy for online data reduction based on reconstruction-based anomaly detection. A convolutional autoencoder trained exclusively on pedestal images (i.e. frames acquired with GEM amplification disabled) learns the detector noise morphology and highlights particle-induced structures through localized reconstruction residuals, from which compact Regions of Interest (ROIs) are extracted. On real prototype data, the selected configuration retains $(93.0 \pm 0.2)\%$ of reconstructed signal intensity while discarding $(97.8 \pm 0.1)\%$ of the image area, with $\sim 25$~ms per-frame inference time on a consumer GPU.

Second, we report a weakly supervised application of the Classification Without Labels (CWoLa) framework to data acquired with an Americium--Beryllium neutron source. Using only mixed AmBe and standard datasets (no event-level labels), a convolutional classifier learns to identify nuclear-recoil-like topologies. The achieved performance approaches the theoretical limit imposed by the mixture composition and isolates a high-score population with compact, approximately circular morphologies consistent with nuclear recoils.
}

\FullConference{
14th Young Researcher Meeting (YRM 2025)\\
29 September -- 3 October 2025\\
Gran Sasso Science Institute (GSSI), L'Aquila, Italy
}

\begin{document}
\maketitle

\section{Introduction}

Optical-readout Time Projection Chambers (TPCs) are powerful detectors for rare-event searches in the ${\cal O}(1$--$100)$~keV regime, where short and localized nuclear-recoil (NR) tracks, the signature of dark-matter interactions, must be identified in the presence of abundant electronic-recoil (ER) backgrounds. In the CYGNO experiment~\cite{Amaro:2022gub}, ionization electrons produced in a He--CF$_4$ gas mixture drift toward a triple Gas Electron Multiplier (GEM) stack, where they undergo charge amplification and induce CF$_4$ electroluminescence. This scintillation light is recorded by scientific CMOS cameras, while photomultiplier tubes provide timing information for three-dimensional reconstruction~\cite{Amaro:2025ssv}. The optical readout offers excellent granularity and sensitivity to low-energy deposits~\cite{Amaro:2023dxb}.

A central challenge is data volume: each exposure is a megapixel-scale image in which the physical signal typically occupies only a small fraction of pixels. This becomes critical for the forthcoming CYGNO-04 demonstrator, where multiple high-resolution cameras at few-Hz rates yield sustained data throughput of $\mathcal{O}(10^2)\,\mathrm{MB/s}$ if full frames are stored. Traditional offline reconstruction pipelines~\cite{Amaro_2023} provide high-fidelity track characterization but are not suitable for trigger-level use, with per-frame latencies of order seconds. CYGNO therefore benefits from (i) \emph{online localization and data reduction} via ROIs, and (ii) \emph{NR/ER discrimination} even when event-level NR labels are not available.

In this contribution we focus on two ML strategies addressing these needs with minimal supervision: (1) a pedestal-trained autoencoder for unsupervised anomaly detection and fast ROI extraction, and (2) a weakly supervised CWoLa analysis using AmBe neutron calibration data to learn NR-like topologies from mixed samples.

\section{Unsupervised ROI extraction from pedestal-trained anomaly detection}
\label{sec:roi}

\paragraph{Pedestal data and preprocessing}

CYGNO camera frames are large, sparse, and dominated by structured noise (readout noise, fixed-pattern features, dark counts, and static optical artifacts). A distinctive asset of this detector is the availability of \emph{pedestal frames} acquired with GEM amplification disabled, providing a high-purity sample of noise-only data for fully unsupervised training.

In this study, pedestal frames (training) are acquired with GEM voltages disabled, while standard frames (evaluation) are taken at nominal gain. To suppress edge instabilities observed in raw images~\cite{Almeida:2025pub}, each frame is fiducialized by cropping to a central $1525\times1525$ region. Inputs are expressed relative to pedestal statistics using a consistent normalization chain and are downscaled to $1024\times1024$ for computational efficiency.

\paragraph{Autoencoder anomaly map and ROI construction}

Reconstruction-based anomaly detection with autoencoders (AEs)~\cite{Goodfellow-et-al-2016,varen_imp,varen5} is well matched to this setting. An AE trained exclusively on pedestal images learns to reconstruct the detector noise, while particle-induced structures present in standard frames appear as localized reconstruction failures. We study a convolutional AE with strided-convolution down-sampling, transposed-convolution up-sampling, skip connections between matching scales, and a 128-dimensional latent representation. For general background on these deep-learning components, see Ref.~\cite{Goodfellow-et-al-2016}. The anomaly map is defined as the pixel-wise residual
$\mathbf{r}(\mathbf{x}) = |\mathbf{x}-\hat{\mathbf{x}}|$.

A baseline configuration employs a hybrid Mean Square Error (MSE) plus Structural Similarity (SSIM) reconstruction objective~\cite{1284395}. Beyond this baseline, alternative training strategies are being studied to enhance sensitivity to faint, localized deviations from pedestal noise and improve anomaly-map quality, while limiting reconstruction of subtle structured features and preserving the learned noise morphology.

Residual maps are converted into ROIs using standard image-processing steps. A global threshold is applied (a representative value $\tau=0.04$, chosen to suppress residual activity on held-out pedestal frames), followed by morphological closing to connect nearby residual fragments along individual tracks (using a circular structuring element with $d_{\rm link}=40$ pixels). The resulting binary mask defines the ROI. The selected working point lies on a broad plateau of the coverage–compression trade-off from a residual-threshold sweep, indicating stability against moderate threshold variations.

\paragraph{Performance}

The extracted ROIs are evaluated against the established offline reconstruction~\cite{Amaro_2023}, which provides a high-fidelity physics reference but is not suitable for online use. Evaluation is performed on a per-event basis, requiring well-contained signals (at least one reconstructed pixel located more than 50 pixels from the image border). We report the signal-intensity coverage, the area cut defined as $f_{\rm cut}=1-A_{\rm ROI}/A_{\rm img}$, and the per-frame inference latency.

On a test sample of 1563 reconstructed events, the current best-performing configuration achieves a signal-intensity coverage of $(93.0\pm0.2)\%$, an area cut of $f_{\rm cut}=(97.8\pm0.1)\%$, and an inference latency of $t_{\rm inf}\simeq 25$~ms per frame. These results illustrate the potential of pedestal-trained anomaly detection as a viable strategy for fast, unsupervised ROI extraction in CYGNO.

A broader discussion of this exploratory approach and of ongoing developments is presented in Ref.~\cite{Amaro:2025jic}, currently submitted for publication to \emph{Machine Learning: Science and Technology} (IOP) and available as an arXiv preprint.
\section{Weakly supervised nuclear-recoil identification with CWoLa}
\label{sec:cwola}

\paragraph{AmBe mixed samples and mixture ceiling}

To obtain an NR-enriched sample without event-level labels, we exploit an Americium--Beryllium (AmBe) neutron source~\cite{ambe}. Neutrons are produced via the reactions
$^{241}\mathrm{Am}\to{}^{237}\mathrm{Np}+\alpha$ and
$^{9}\mathrm{Be}+\alpha\to{}^{12}\mathrm{C}^*+n$,
with frequent de-excitation of ${}^{12}\mathrm{C}^*$ emitting a 4.43~MeV $\gamma$ ray. The emitted neutron spectrum extends up to $\sim 11$~MeV with a mean of $\sim 4$--$5$~MeV~\cite{ambe}. A dedicated GEANT4 study for a CYGNO-like configuration~\cite{DiGiambattista2024} indicates that neutrons contribute significantly to the AmBe-induced response in the active volume, producing NRs directly and additional ERs via associated $\gamma$ production in surrounding materials.
We define two datasets: an \emph{AmBe} sample (source present, signal-enriched) and a \emph{standard} (STD) sample (source absent, background-dominated), both restricted to a common region of the reconstructed energy--density plane (where the density is defined as the reconstructed energy divided by the number of hit pixels) in which AmBe shows a clear rate excess. In this region, a data-driven rate comparison yields a signal-attributable fraction $\alpha=(32.0\pm0.9)\%$. This mixture structure imposes a strict theoretical ceiling on ROC--AUC,
$\mathrm{AUC}_{\max}=0.5+\alpha/2=(0.660\pm0.005)$, since the irreducible $(1-\alpha)$
background component in the AmBe data is indistinguishable from STD~\cite{Metodiev:2017vrx}.

\paragraph{Classifier and results}

\begin{figure}[t]
\centering
\includegraphics[width=0.6\textwidth]{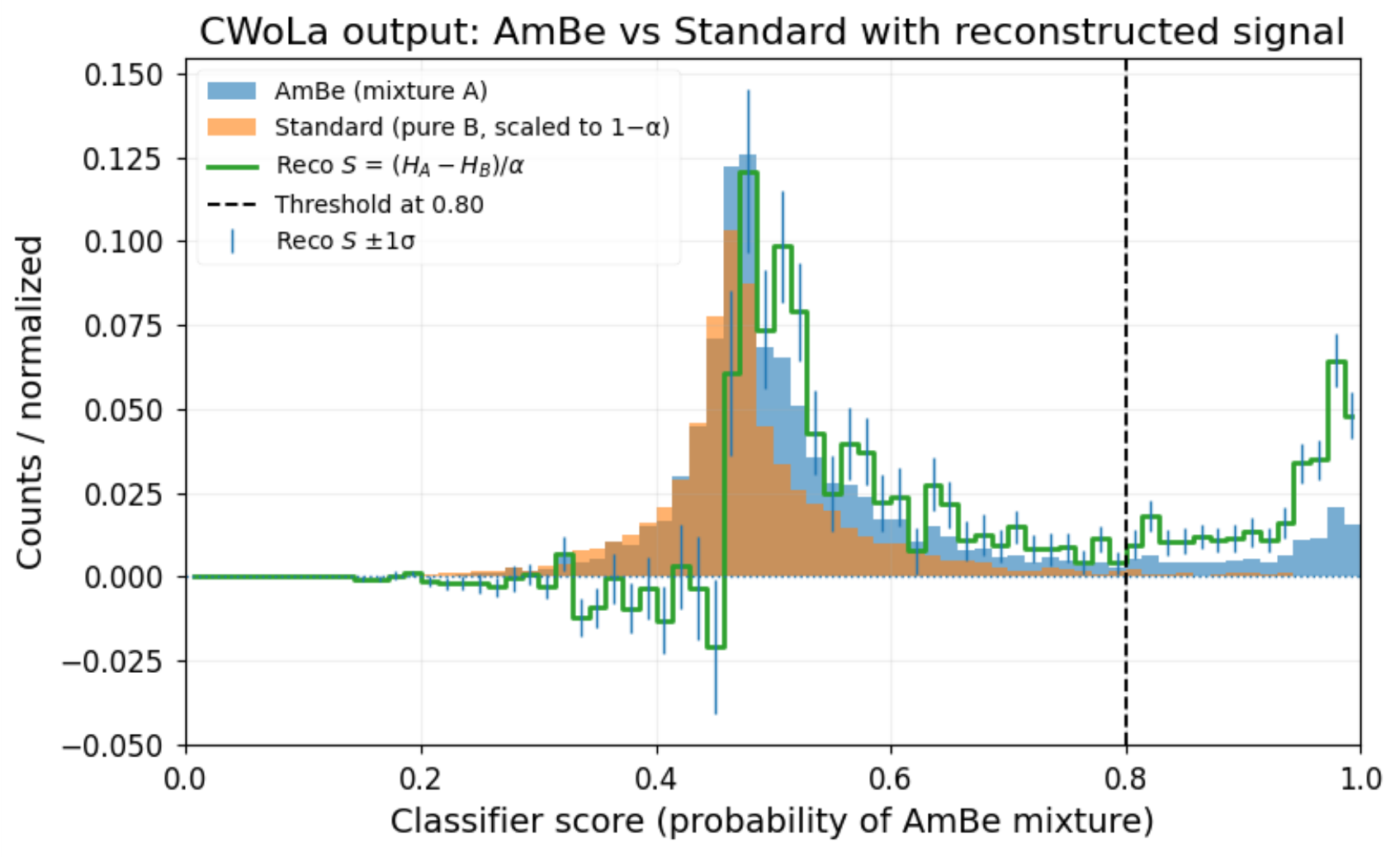}
\caption{CWoLa classifier-score distributions for AmBe and standard test samples. The reconstructed signal component exhibits a peak near unity (NR-dominated) and an additional enhancement around $\sim 0.5$ (overlapping AmBe-induced populations). The dashed line indicates the NR-selection threshold.}
\label{fig:cwola_scores_pos}
\end{figure}

The CWoLa approach~\cite{Metodiev:2017vrx} trains a classifier to distinguish the two mixtures using only sample-level labels; asymptotically this recovers the optimal $S$ vs.\ $B$ discriminator. We train a compact convolutional neural network on $128\times128$ grayscale images of reconstructed events, using the Adam optimizer~\cite{Kingma2014AdamAM} and binary cross-entropy while monitoring AUC. The test-set score distributions in Fig.~\ref{fig:cwola_scores_pos} show an excess of high-score events in AmBe relative to STD, with a peak near unity (NR-dominated) and an enhancement around $\sim0.5$ (overlapping source-induced populations). We select the NR-dominated tail with a conservative threshold $p>0.8$.

Events above threshold populate the NR-favored region of the energy–density plane and show compact, approximately circular morphologies with high reconstructed density, consistent with nuclear recoils. These features serve as a qualitative consistency check, while quantitative validation with labeled simulations is deferred to future studies. The achieved performance approaches the theoretical ceiling, indicating near-optimal extraction of separation power from the mixed samples.

\section{Conclusions}
We summarized two complementary ML approaches for CYGNO with minimal supervision. A pedestal-trained anomaly-detection pipeline enables fast ROI selection from sparsely populated megapixel images, retaining $(93.0\pm0.2)\%$ of reconstructed signal intensity while discarding $(97.8\pm0.1)\%$ of image area at $\sim25$~ms/frame, providing a realistic baseline for online data reduction. In parallel, a weakly supervised CWoLa classifier trained on mixed AmBe and standard samples learns NR-like topologies directly from data. A data-driven signal fraction $\alpha=(32.0\pm0.9)\%$ sets a theoretical AUC ceiling of $(0.660\pm0.005)$, and the observed performance approaches this limit while isolating compact, near-circular NR-like events. Together, these results support an ML-assisted path toward scalable online selection in next-generation CYGNO optical TPCs and, at the level of underlying methodology, more broadly applicable strategies for rare-signal localization and weakly supervised learning in other dark-matter searches.

\paragraph{Acknowledgements}

This project has received fundings under the European Union’s Horizon 2020 research and innovation program from the European Research Council (ERC) grant agreement No. 818744 and is supported by the Italian Ministry of Education, University and Research through the project PRIN: Progetti di Ricerca di Rilevante Interesse Nazionale “Zero Radioactivity in Future experiment” (Prot. 2017T54J9J). A. Messina has also been supported by the PNRR MUR project PE0000013–FAIR. 


\bibliographystyle{unsrt}
\setlength{\bibsep}{0pt}
\bibliography{cygno_pos}

\end{document}